\documentclass{aa}

\usepackage{txfonts}
\usepackage{graphicx}
\usepackage{natbib}

\usepackage{soul}

\newcommand{\ks}{$K_{S}$}

\newcommand{\lmxb}{LMXB}
\newcommand{\lmxbs}{LMXBs}
\newcommand{\psf}{{\it PSF}}
\newcommand{\psfs}{{\it PSFs}}

\newcommand{\xtefifteen}{XTE\,J1550$-$564}

\begin{document}


\title{NIR and optical observations of the failed outbursts of\\ black
  hole binary \xtefifteen\thanks{Based on observations made with the
    European Southern Observatory telescopes obtained from the
    ESO/ST-ECF Science Archive Facility.}  }


\titlerunning{NIR and optical observations of \xtefifteen}

\author{
  P.A.~Curran\inst{1,2}
  \and S.~Chaty\inst{2,3}
}

\institute{International Centre for Radio Astronomy Research, Curtin
  University, GPO Box U1987, Perth, WA 6845, Australia
  \and  AIM (UMR-E 9005 CEA/DSM-CNRS-Universit\'e Paris Diderot), Irfu/Service
 d'Astrophysique, Centre de Saclay, FR-91191 Gif-sur-Yvette Cedex,
 France
\and Institut Universitaire de France, 103, boulevard Saint-Michel, 75005
 Paris, France
}

\date{Received 9 May 2013 / Accepted 17 July 2013}


\abstract
{ A number of low mass X-ray binaries (\lmxbs) undergo ``failed outbursts'' in
  which, instead of evolving through the canonical states, they remain
  in a hard state throughout the outburst. While the sources of X-ray
  and radio emission in the hard state are relatively well understood,
  the origin of the near infrared (NIR) and optical emission is more
  complex though it likely stems from an amalgam of different
  emission processes, occurring as it does, at the intersecting
  wavelengths of those processes.}
%
{ We aim to identify the NIR/optical emission region(s) during a
  number of failed outbursts of one such low mass X-ray binary and
  black hole candidate, \xtefifteen, in order to confirm or refute
  their classification as hard-state, failed outbursts.}
%
{ We present unique NIR/optical images and spectra, obtained with the
  ESO--New Technology Telescope, during the failed outbursts of 2001
  and 2000. We compare the NIR/optical photometric, timing, and spectral
  properties with those expected for the different emission mechanisms
  in the various  \lmxb\ states.  }
%
{The NIR/optical data are consistent with having come from
  reprocessing of X-rays in the accretion disk, with no evidence of
  direct thermal emission from the disk itself. However, the observed
  variability in high-cadence NIR light curves suggest that the radio
  jet extends and contributes to the NIR wavelengths.}
%
{We find that these failed outbursts did not transition to an
  intermediate state but remained in a true, hard state where there
  was no sign of jet quenching or deviation from the observed hard
  state correlations.}


\maketitle

\section{Introduction}\label{section:intro}

 \begin{figure*}
  \centering 
  \resizebox{\hsize}{!}{\includegraphics[angle=-90]{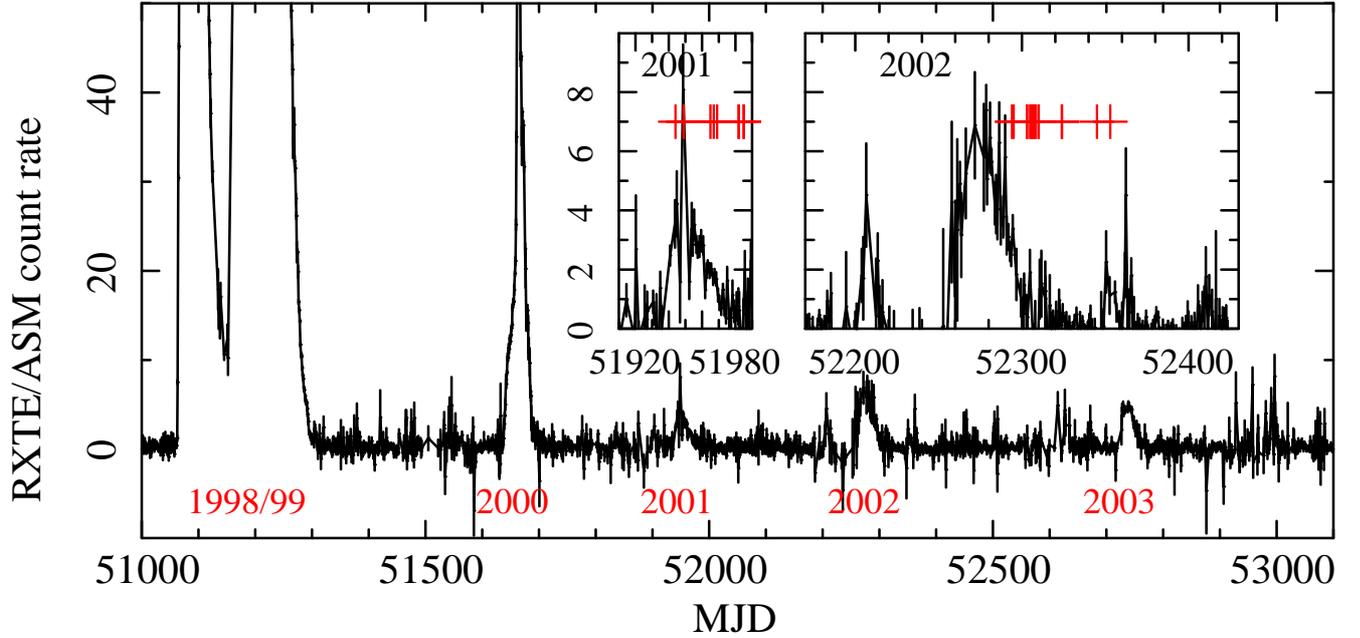}}
  \caption{The one-day averaged RXTE/ASM light curve for the period in
    question (count rate axes cut off at 50 counts/second for clarity
    but peaked at $\approx 490$ counts/second during 1998
    outburst). The epochs of our observations are marked with crosses
    on the zoomed inlays for the failed outbursts of 2001 and 2002.}
  \label{fig.rxte} 
\end{figure*}

For the majority of their lifetimes, transient low mass
X-ray binaries (\lmxbs) are in a state of quiescence with faint or
non-detected X-ray emission. In quiescence, near infrared (NIR) and
optical emission, if detected, is dominated by the main-sequence
companion star (with possibly significant contribution from the cold
accretion disk). During outburst -- on time scales of weeks, months
or even longer -- there is a dramatic increase in the X-ray,
NIR/optical, and radio flux, which is powered by an increased level of
accretion onto the central, compact object (black hole or neutron
star). Many of these sources are observed to undergo multiple,
irregular outbursts (e.g., \xtefifteen\ has displayed 5 outburst
events in less than a decade) while others may remain in quiescence
for decades after their initial discovery (e.g., V2107\,Oph was
detected in outburst in 1977 but has yet to repeat).

Black hole \lmxb\ outbursts are usually divided into a number of {\it
  states}, based mainly on observable X-ray spectral and timing
characteristics.  The sources are initially observed in a, generally
low intensity, {\it hard} state with spectra dominated by power-law
emission. They then transition, via an {\it intermediate} state, to a
{\it soft} or {\it thermal-dominant} state (so called because the
spectrum is dominated by a thermal component). X-ray flux peaks in
this state before decreasing and evolving, via a late hard state, back
into a quiescent state. For a fuller description of the various
possible states and the associated X-ray timing properties, etc., see
\citealt{mclintock2006:csxs157}.
While the majority of outbursts from \lmxbs\ follow this standard
evolution of X-ray defined states, a number of sources are observed to
return to quiescence without displaying a soft state (e.g., 9 sources
in \citealt{Brocksopp2004:NewA.9} and references therein) and another
four have been observed to proceed to an intermediate state before
returning to the hard state and quiescence, without reaching the soft
state
\citep{Capitanio2009:MNRAS.398,Ferrigno2012:A&A.537,Soleri2013:MNRAS.429,Wijnands2002:ApJ.564}.
Despite being referred to as ``failed outbursts'' (or {\it ``Soft
  X-ray transient'' outbursts which are not soft};
\citealt{Brocksopp2004:NewA.9}) these outbursts can in fact be quite
luminous (e.g., V404 Cyg; \citealt{Tanaka1995xrbi.nasa}), though most
are under-luminous.

While the origin of X-ray emission in the different states is
relatively well understood (e.g., \citealt{mclintock2006:csxs157}),
that of the NIR, optical, and ultraviolet (UV) is more complex as the
optical wavelengths are at the intersection of a number of different
emission mechanisms (for reviews of optical properties of \lmxbs\ see
e.g., \citealt{vanParadijs1995:xrbi.nasa.58,Charles2006:csxs.book}).
Both intrinsic, thermal emission from the hot, outer accretion disk
(e.g., \citealt{Shakura1973:A&A.24,Frank2002:apa.book}) as well as
reprocessing of X-rays in the same region of the disk (e.g.,
\citealt{Cunningham1976:ApJ.208,vanParadijs1994:A&A.290}) may
contribute significant levels of flux at UV, optical, and NIR
wavelengths.  Recently, evidence has been mounting that the
relativistic jet, usually detected in radio, also produces a
significant contribution to the NIR -- and possibly optical -- flux,
at least in the hard state (e.g.,
\citealt{Jain2001:ApJ554,Corbel2002:ApJ.573,Russell2006:MNRAS.371,Chaty2011:A&A.529}),
and it is possible that the power law component of the X-ray emission
extends to, and contributes at optical wavelengths.

\xtefifteen\ has undergone a number of weak,
failed outbursts (in 2001 \citep{Tomsick2001:IAUC.7575}; 2002
\citep{Belloni2002:A&A.390}; and 2003
\citep{Sturner2005:ApJ.625,Arefev2004:AstL.30}) as well as a number
of complete (soft state) outbursts in 1998/99 (e.g.,
\citealt{Sobczak2000:ApJ.544}) and 2000 (e.g.,
\citealt{Jain2001:ApJ554}; see also figure \ref{fig.rxte}).  Even if
these late time outbursts are considered as rebrightenings or
reflarings of the original outburst, this source demonstrates
conclusively that failed outbursts are not a separate class of object
but are likely caused by differences in the accretion flow onto the
black hole or by differences in the systems' efficiency in converting
the accreted matter into observable flux.
In this paper we present the only significant NIR/optical observations
during the failed outbursts of 2001 and 2002, obtained by the ESO NTT
(Table\,\ref{table:nights}), and comprising of all the available
unpublished, archived ESO data of the source.  In section
\ref{section:observations} we introduce the observations and reduction
methods, while in section \ref{section:results} we present the results
of our photometric, timing and spectral analyses of the data. We
discuss the interpretation of our findings in section
\ref{section:discussion} and summarise in section
\ref{section:conclusions}.


\section{Observations \& Reduction}\label{section:observations}

\subsection{Photometry}

\begin{table}	
  \centering	
  \caption{Nights of observations.} 	
  \label{table:nights} 	
  \begin{tabular}{l l l l l} 
    \hline\hline
    Night  & MJD & Filters  & Size($\arcmin$) & ESO ID\\ 
    \hline 
    February 2 2001       & 51943 & $J,H,K_{S}$ & $4\times4$  & 66.C-0120\\
    February 7 2001       & 51948 & $J,H,K_{S}$ & $4\times4$  & 59.A-9004\\
    & 51948 & $GRF$* & --- & 59.A-9004   \\
    February 23 2001     & 51964 & $V,R,I$    & $8\times9$  & 59.A-9907\\
    February 26 2001     & 51966 & $RILD\#1$*  & ---  & 66.D-0199 \\
    February 27 2001     & 51968 & $V,R,I$    & $8\times9$  & 66.A-0617\\
    March  12 2001       & 51981 & $K_{S}$     & $4\times4$  & 66.A-0162\\
    March  15 2001       & 51984 & $K_{S}$     & $4\times4$  & 66.A-0162\\

    January 18 2002     & 52293 & $V,R,I$$^{s}$    & $5\times5$  & 59.A-9004\\
    January 19 2002     & 52294 & $J,H,K_{S}$ & $4\times4$  & 59.A-9004\\
    January 27 2002     & 52302 & $V$         & $2\times2$  & 68.D-0316\\
    & 52302 & $RILD\#1$*     & ---  & 68.D-0316 \\
    January 29 2002     & 52304 & $K_{S}$ & $4\times4$  & 68.D-0316\\
    January 30 2002     & 52305 & $GBF,GRF$*& ---  & 68.D-0316 \\
    January 31 2002     & 52306 & $J,H,K_{S};K_{S}$ & $4\times4$  & 68.D-0316\\
    February 1 2002     & 52307 & $GBF,GRF$*& ---  & 68.D-0316 \\
    February 3 2002     & 52309 & $V,R,I$    & $8\times8$  & 68.D-0316\\
    & 52309 & $RILD\#1$* & ---  & 68.D-0316 \\
    February 17 2002   & 52323 & $V$    & $2\times2$  & 68.D-0316\\
    March 10 2002     & 52344 & $V,R$$^{f}$    & $7\times7$  & 68.D-5771\\
    March 18 2002     & 52352 & $I$$^{f}$    & $7\times7$  & 68.D-5771\\
    \hline 
  \end{tabular}
  \begin{list}{}{} 
  \item[]{\bf Notes.}* Spectra; All NIR data were obtained by SofI while optical
    data were obtained by EMMI unless otherwise noted as $^{s}$ (SUSI2)
    or  $^{f}$ (FORS1).
  \end{list}
\end{table}

During the outbursts of 2001 and 2002 optical ($V,R,I$) data were
obtained with the ESO Multi-Mode Instrument (EMMI;
\citealt{Dekker1986:SPIE627}) and the Superb-Seeing Imager (SuSI2) on
the $3.58$m ESO -- New Technology Telescope (NTT), as well as with the
FOcal Reducer and low dispersion Spectrograph (FORS1) on the $8.2$m
UNIT 3 of the Very Large Telescope (VLT-UT3)
(Table\,\ref{table:nights}). Furthermore, during the same period NIR
($J,H,K_{S}$) data were obtained with the Son of ISAAC (SofI) infrared
spectrograph and imaging camera on the NTT. 
These data were reduced using the {\small IRAF} package wherein
crosstalk correction, bias-subtraction, flatfielding, sky subtraction,
bad pixel correction and frame addition were carried out as
necessary. The dither pattern, necessary for sky subtraction, was not
applied to some of the $J, H$, and \ks\ images obtained on MJD 52306 so
those images underwent no further analysis.

\begin{figure} 
  \centering 
  \resizebox{\hsize}{!}{\includegraphics[angle=-0]{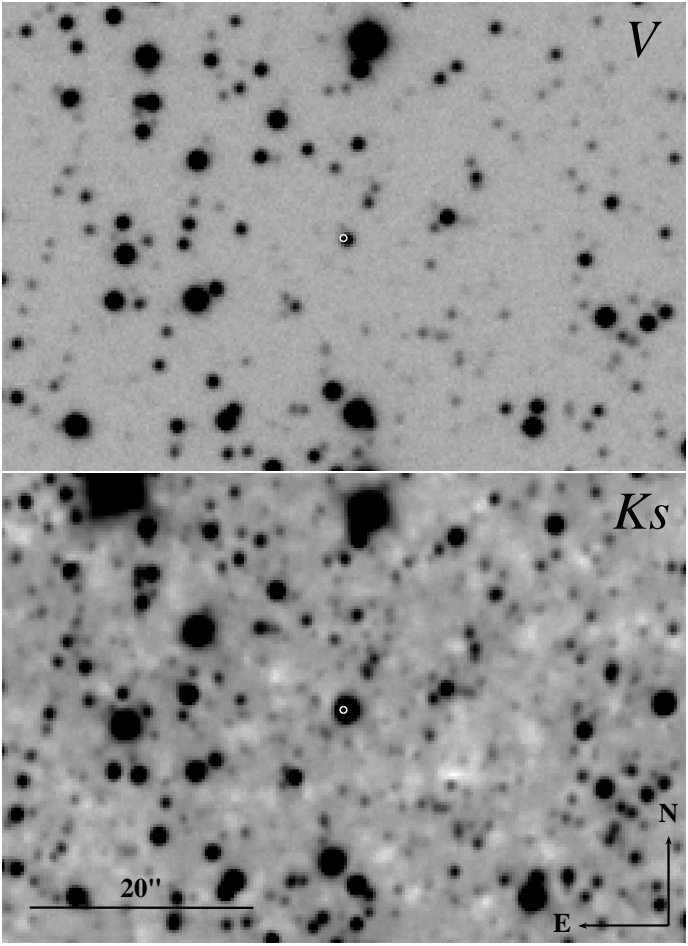}}
  \caption{NTT 60\arcsec\ $\times$ 40\arcsec\ finding charts
    ({\it upper:} 530s $V$ image on MJD 52324; {\it lower:} 3240s \ks\ image on MJD
    52305) with the 0.3\arcsec\ radio positional uncertainty 
    \citep{Corbel2001:ApJ.554} marked by a circle. }
  \label{fig.finding_chart} 
\end{figure}

The images were astrometrically calibrated within the GAIA package,
against the 2MASS \citep{Skrutskie2006:AJ.131} or USNO-B1.0
\citep{Monet2003:AJ125} catalogues.  The position of \xtefifteen\ was
derived via the point spread function (\psf) of the source in the deep
(3240s) \ks-band image on January 30 (MJD 52305; seeing $\approx
0.8$\arcsec) as 15:50:58.67 $-$56:28:35.3, with a positional
error\footnote{All uncertainties in this paper are given with a
  confidence of $1\sigma$.} dominated by the 0.1\arcsec\ 2MASS
systematic uncertainty (Figure\,\ref{fig.finding_chart}). This is
consistent with the radio position of \cite{Corbel2001:ApJ.554} and
the optical position of \cite{Jain1999:ApJ.517L}.

\begin{figure} 
  \centering 
  \resizebox{\hsize}{!}{\includegraphics[angle=-90]{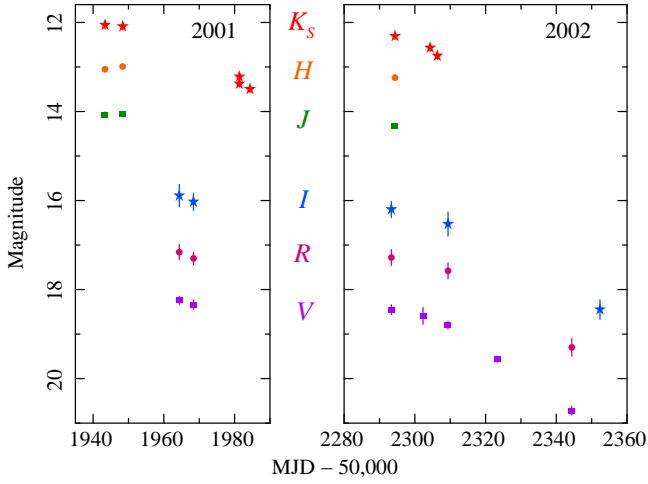}}
  \caption{Optical and near-infrared light curves of \xtefifteen\
    during the failed outbursts of 2001 and 2002.}
  \label{fig.lightcurve}
\end{figure}

Relative \psf\ photometry was carried out on the final images using
the {\small DAOPHOT} package \citep{stetson1987:PASP99} within {\small
  IRAF}.  The NIR and $V$-band magnitudes
(Table\,\ref{table:magnitudes}, Figure\,\ref{fig.lightcurve}) were
calibrated against the 2MASS and GSC 2.3 \citep{Russell1990:AJ.99}
catalogues using $\approx 100 - 300$ objects per image, after outliers
and saturated objects were removed. $R$- and $I$-band magnitudes were
calibrated against field stars observed by \cite{Jain2001:ApJ546} and
\cite{Sanchez1999:A&A.348}, noting that the published positions of the
latter are incorrect and using the transformation, $i-I = (0.247 \pm
0.003) (R-I)$ \citep{Jordi2006A&A.460}. $V$-band magnitudes were also
estimated by this method, and were consistent with the GSC derived
values. All derived magnitudes were comparable to magnitudes estimated
via \cite{Persson1998:AJ116} and \cite{Landolt1992:AJ.104} photometric
standards observed on some of the nights.  Due to the small field of
view ($2\arcmin\times2\arcmin$) some $V$-band images, only $\approx
20$ objects per image were available in these cases. The FORS1 images
were heavily affected by saturation of catalogue sources in the field
so magnitudes were derived from 10 relatively isolated field stars,
which were in turn derived from the catalogues on nights less affected
by saturation.


On a number of nights, data consisting of multiple ($\geq 53$)
high-cadence, ``fast'' photometry, images in \ks- and $V$-band were
obtained in order to investigate possible short-term variability of
the source (see table \ref{table:magnitudes}).  \psf\ photometry was
carried out on each of these individual images, again using {\small
  DAOPHOT}. In each, the source magnitude was calculated relative to a
number of field stars (12 in NIR and 10 in optical) and normalised so
that the average is equal to zero. In addition, we also calculate
(relative to the same field stars but normalised to a magnitude of
one) the magnitudes of 5 comparison stars per band, of similar
magnitude to \xtefifteen\ (Table\,\ref{table:comparisons}).  The
positions of those objects are derived via \psfs\ in the 530s $V$-band
image on MJD 52324 or the 3240s \ks-band image on MJD 52305
(Figure\,\ref{fig.finding_chart}; seeing of both $\approx 0.8$\arcsec)
and are dominated by the 0.1\arcsec\ 2MASS systematic uncertainty.

\begin{table}	
  \centering	
  \caption{Optical and NIR exposures and magnitudes.} 	
  \label{table:magnitudes} 	
  \begin{tabular}{l l l l} 
    \hline\hline
    MJD & Filter & N $\times$ Exp (s) & Magnitude  \\ 
    \hline    
    51964.36639 & $V$  & $1\times300$  &   18.24 $\pm$ 0.09 \\
    51968.39107 & $V$  & $1\times300$  &   18.34 $\pm$ 0.11 \\
    52293.35916 & $V$  & $1\times300$  &   18.45 $\pm$ 0.11 \\
    52302.31761 & $V$  & $81\times10$  &   18.59 $\pm$ 0.19 \\
    52309.36426 & $V$  & $1\times300$  &   18.80 $\pm$ 0.09 \\
    52323.35353 & $V$  & $53\times10$  &   19.57 $\pm$ 0.06 \\
    52344.37184 & $V$  & $2\times300$  &   20.72 $\pm$ 0.10 \\
    51964.37249 & $R$  & $1\times300$  &   17.16 $\pm$ 0.17 \\
    51968.38505 & $R$  & $1\times300$  &   17.30 $\pm$ 0.14 \\
    52293.36300 & $R$  & $1\times300$  &   17.28 $\pm$ 0.18 \\
    52309.37640 & $R$  & $1\times300$  &   17.58 $\pm$ 0.18 \\
    52344.38006 & $R$  & $2\times400$  &   19.30 $\pm$ 0.20 \\
    51964.37850 & $I$  & $1\times300$  &   15.89 $\pm$ 0.25 \\
    51968.37908 & $I$  & $1\times300$  &   16.03 $\pm$ 0.19 \\
    52293.36683 & $I$  & $1\times300$  &   16.20 $\pm$ 0.18 \\
    52309.38240 & $I$  & $1\times300$  &   16.53 $\pm$ 0.27  \\
    52352.34852 & $I$  & $5\times240$  &   18.45 $\pm$ 0.22 \\
    51943.35829 & $J$ & $18\times10$   &  14.07 $\pm$ 0.06 \\
    51948.35504 & $J$ & $9\times10$    &  14.06 $\pm$ 0.04 \\
    52294.37304 & $J$ & $7\times10$    &  14.32 $\pm$ 0.05  \\
    %
    51943.37134 & $H$ & $9\times10$    &  13.05 $\pm$ 0.06 \\
    51948.36251 & $H$ & $9\times10$    &  12.99 $\pm$ 0.06 \\
    52294.37886 & $H$ & $7\times10$    &  13.24 $\pm$ 0.06 \\
    %
    51943.37880 & \ks & $9\times10$ &  12.06 $\pm$ 0.09 \\
    51948.37000 & \ks & $9\times10$ &  12.09 $\pm$ 0.07 \\
    51981.34750 & \ks & $54\times5$ &  13.38 $\pm$ 0.07 \\
    51981.35707 & \ks & $270\times2$ & 13.22 $\pm$ 0.07 \\
    51984.34401 & \ks & $324\times2$ & 13.50 $\pm$ 0.07 \\
    52294.38468 & \ks & $7\times10$ &  12.31 $\pm$ 0.07  \\
    52304.31679 & \ks& $324\times10$ & 12.57 $\pm$ 0.08  \\
    52306.34938 & \ks& $243\times10$ & 12.75 $\pm$ 0.08 \\
  \hline 
    51948.385  & $GRF$ & $10\times120$  & ---\\
    51966.363  & $RILD\#1$ & $6\times300$ &  ---\\
    52302.290  & $RILD\#1$ & $3\times600$ &  ---\\
    52305.359  & $GBF$ & $8\times240$  & --- \\
    52305.385  & $GRF$ & $8\times240$  & ---\\
    52307.335  & $GBF$ & $8\times240$  & --- \\
    52307.358  & $GRF$ & $8\times240$  & ---\\
    52309.339  & $RILD\#1$ & $3\times600$ &  ---\\
\hline
\end{tabular}
\end{table}

\begin{table}	
  \centering	
  \caption{Positions and magnitudes of sources used for comparison
    against high-cadence photometry.}
  \label{table:comparisons} 	

  \begin{tabular}{l l l} 
    \hline\hline
       & Position & Magnitude \\ 
    \hline     
    $V$  & &    \\
    \hline     
       & 15:50:58.19 $-$56:27:58.2 &  19.35 $\pm$ 0.19 \\
       & 15:51:00.28 $-$56:29:03.0 &  18.89 $\pm$ 0.19 \\
       & 15:50:55.82 $-$56:28:32.0 &  19.83 $\pm$ 0.19 \\
       & 15:50:56.72 $-$56:29:15.8 &  19.03 $\pm$ 0.19 \\
       & 15:50:57.93 $-$56:29:00.3 &  19.54 $\pm$ 0.14 \\
    \hline     
    \ks  & &    \\
    \hline     
      & 15:50:58.41 $-$56:28:11.3 & 12.595 $\pm$ 0.038 \\
      & 15:50:58.82 $-$56:28:48.9 & 12.379 $\pm$ 0.040\\
      & 15:50:55.25 $-$56:28:34.7 & 12.969 $\pm$ 0.035 \\
      & 15:50:56.73 $-$56:29:15.7 & 13.563 $\pm$ 0.050  \\
      & 15:50:57.56 $-$56:29:12.1 & 12.721 $\pm$ 0.033   \\
\hline
\end{tabular}
 \begin{list}{}{}
 \item[]{\bf Notes.} $V$-band magnitudes are derived
   as described in section \ref{section:observations}, while \ks
   magnitudes are extracted directly from the 2MASS catalogue.
 \end{list}
\end{table}

\subsection{Spectral Energy Distributions}

 \begin{figure} 
  \centering 
  \resizebox{\hsize}{!}{\includegraphics[angle=0]{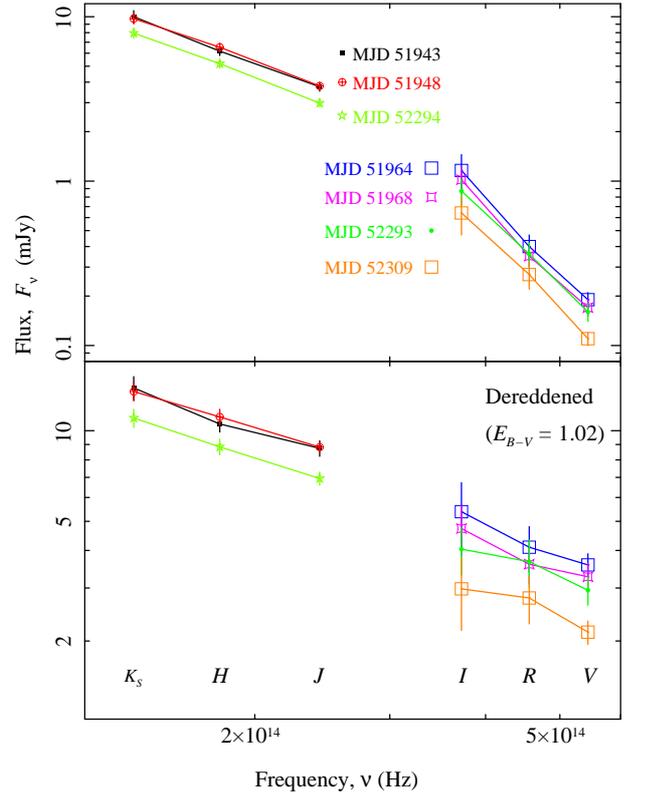}}
  \caption{Flux versus frequency plot, at 6 different epochs,
    uncorrected (upper) and corrected (lower; \citealt{cardelli1989:ApJ345}) for the derived
    extinction in the direction of the source of $E_{B-V} = 1.02 \pm 0.05$.}
  \label{fig.OIR_SED} 
\end{figure}

For the purposes of fitting, the observed magnitudes
(Table\,\ref{table:magnitudes}) were converted to flux densities,
$F_{\nu}$, at frequency $\nu$ (Figure\,\ref{fig.OIR_SED}), and then to
flux per filter, $F_{filter}$ in units of
photons\,cm$^{-2}$\,s$^{-1}$.  This is done via $F_{filter} =
1509.18896 F_{\nu}$ $( \Delta\lambda/\lambda )$ where $\lambda$ and
$\Delta\lambda$ are the effective wavelength and full width at half
maximum of the filter in question.  {\tt XSPEC} compatible files, for
Spectral Energy Distribution (SED) fitting, were produced from the
flux per filter value using the {\tt FTOOL}, {\tt flx2xsp}. Due to the
time difference between epochs of observations, we treat each night
separately except for the data taken on the adjoining nights of MJD
52293 and MJD 52294. Due to the time difference between the final
I-band image (on MJD 52352) and the corresponding $V$- and $R$-band
images (on MJD 52344) we do not consider these in our analysis, this
results in 6 independent epochs (see table \ref{table:x-ray}).

Our NIR/optical data were augmented with X-ray spectral data from the
Rossi X-ray Timing Explorer (RXTE). Pre-processed Proportional Counter
Array (PCA) and High Energy X-ray Timing Experiment (HEXTE; clusters 0
and 1) ``Standard Product'' spectra were downloaded from the HEASARC
archive for each of the seven nights where we had simultaneous optical
or NIR data. All RXTE spectra were observed within 0.1-1.0 days of our
observations.  PCA spectra were fit from 3-25\,keV while HEXTE were
fit from 25-150\,keV. Unabsorbed X-ray fluxes, from 2-10\,keV, (for
comparison with correlations) were inferred from power-law fits to the
spectra at each epoch (Table\,\ref{table:x-ray}).

\begin{table}	
  \centering  
  \caption{Unabsorbed X-ray fluxes, from 2-10 keV.} 
  \label{table:x-ray} 	
  \begin{tabular}{l l} 
    \hline\hline
    MJD & $F_{{\rm 2-10\,keV}}$ \\ 
    &  ergs\,cm$^{-2}$s$^{-1}$ \\ 
    \hline    
    51942.730 &   7.1$\times 10^{-8}$ \\ 
    51948.501 &   8.5$\times 10^{-8}$ \\ 
    51964.309 &   6.4$\times 10^{-8}$ \\ 
    51968.590 &   4.4$\times 10^{-8}$ \\ 
    52293.447 &   7.8$\times 10^{-8}$ \\ 
    52308.843 &   1.4$\times 10^{-8}$ \\ 
\hline
\end{tabular}
\end{table}

\subsection{Spectroscopy}

Spectral images were obtained on a number of nights
(Tables\,\ref{table:nights}, \ref{table:magnitudes}) with EMMI
obtaining red (3,850--10,000\,\AA), low-dispersion spectra using
Grism\,\#1 ($RILD\,1$) and SofI obtaining blue ($GBF$;
9,500--16,400\,\AA) and red ($GRF$; 15,300--25,200\,\AA) low
resolution spectra.  The data were reduced using the {\small IRAF}
package wherein crosstalk correction, flatfielding, and bias
subtraction were carried out as necessary.  To correct for NIR sky,
the dithered NIR exposures were summed to create sky images which were
subtracted.

Spectra were reduced and extracted within the {\small IRAF}
package, {\tt noao.twodspec}, and individual exposures of the same
spectra were summed.
Due to the crowded nature of the field and a lack of acquisition
frames on some nights we were only able to extract spectra for the
nights of MJDs 52302, 52305, and 52307.
On nights when they were available, wavelength calibrations
were performed against helium + argon (optical) or xenon (NIR) lamps
whose spectra were extracted using the same parameters as for the
relevant source.  The (wavelength dependent) resolution of the final
spectrum is 7--10\,\AA, with a wavelength calibration error of
$\lesssim 20$\,\AA\ (optical) or $\lesssim 40$\,\AA\ (NIR).
Atmospheric, telluric features significantly affect the spectra and
are corrected for by dividing the source spectrum by that of a
telluric standard at a similar airmass, using the {\tt telluric} tool
within {\small IRAF}. This procedure often causes artefacts in the
corrected spectra and in the case of the optical spectra, these
artefacts are dominant so this procedure is not applied. Neither
the optical nor the NIR spectra can be flux calibrated, due to a lack
of standards, but they have been normalised.

No significant features which could not be associated to artefacts of
the telluric correction are found in any of the extracted spectra. To
increase the signal to noise, the NIR spectra from MJDs 52305 and
52307 (which exhibited consistent features) were summed but this did
not exhibit any additional features.


\section{Results}\label{section:results}

\begin{table*}
  \centering	
  \caption{Optical and NIR variability of high-cadence photometry. }
  \label{table:variability} 	

  \begin{tabular}{l l l l l l } 
    \hline\hline
    MJD & Filter & STD & \% RMS & $\chi^{2}_{\nu}$   & dof \\ 
    \hline    
    52302.31761 & $V$  & 0.076 ($\leq0.091$) & $-1.4$ $\pm$ 8.7  ($<26.2$)   &   3.6 ($\leq1.7$) & 80   \\
    52323.35353 & $V$  & 0.108 ($\leq0.038$) & 6.7 $\pm$ 3.6 ($<10.7$)  & 10.4 ($\leq1.6$) & 52   \\
    51981.34750   & \ks  & 0.234 ($\leq0.056$) & 17.8 $\pm$ 5.3 ($<15.9$) &  63.3 ($\leq1.8$) & 323  \\
    51984.34401 & \ks  & 0.274 ($\leq0.069$) & 20.8 $\pm$ 6.6 ($<19.7$) & 68.3 ($\leq2.3$) & 323  \\
    52304.31679 & \ks  & 0.153 ($\leq0.049$) &  10.0 $\pm$ 4.6 ($<13.9$)   &  63.0 ($\leq4.9$) & 323  \\
    52306.34938 & \ks  & 0.221 ($\leq0.056$) & 16.4 $\pm$ 5.3 ($<15.9$)  &  33.7 ($\leq2.0$) & 242  \\
\hline
\end{tabular} 
\begin{list}{}{}
  \item[]{\bf Notes.} The standard deviation of the magnitudes, STD,
    and the associated root mean square variability as percentage of
    the flux, \%RMS, and the $\chi^{2}_{\nu}$ (and number of degrees
    of freedom, dof) of a constant fit to the data are given for the
    source (and, in brakets, for the worst case of the comparison
    sources, or the upper \%RMS limit implied from the comparison
    sources).
  \end{list}
\end{table*}

\subsection{High-cadence photometry}

\begin{figure} 
  \centering 
  \resizebox{\hsize}{!}{\includegraphics[angle=-90]{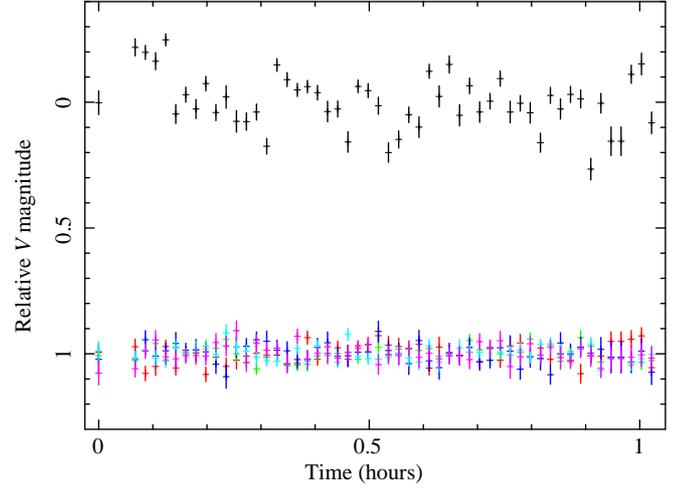}}
  \caption{High-cadence $V$-band light curves of \xtefifteen\
    (normalised to a relative magnitude of 0) and of 5 comparison
    sources (normalised to relative magnitudes of 1) on MJD
    52323.}
  \label{fig.lc_V_fast} 
\end{figure}

\begin{figure} 
  \centering 
  \resizebox{\hsize}{!}{\includegraphics[angle=-90]{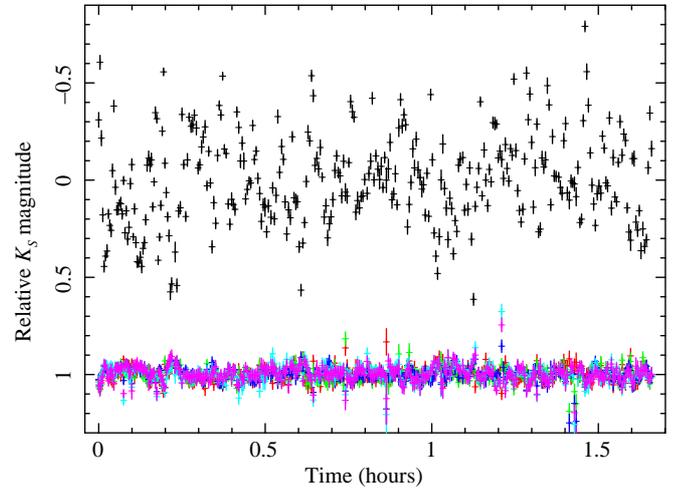}}
  \caption{High-cadence \ks-band light curves of \xtefifteen\
    (normalised to a relative magnitude of 0) and of 5 comparison
    sources (normalised to relative magnitudes of 1) on MJD
    51981.}
  \label{fig.lc_K_fast} 
\end{figure}

The high-cadence $V$- and \ks-band light curves of
\xtefifteen\ exhibit significant variability over the observations
(see Figures \ref{fig.lc_V_fast} and \ref{fig.lc_K_fast} for examples
on MJD 52323 and 51981, respectively), though this is much stronger in
the NIR \ks-band than the optical $V$-band. In all four \ks-band
light curves, the $\chi^{2}_{\nu}$ of a constant fit to the data is
inconsistent ($>5\sigma$) with being acceptable and is significantly
greater than the $\chi^{2}_{\nu}$ of a constant fit to the any of the
five comparison objects of similar magnitude
(Table\,\ref{table:comparisons}), even in the worst case of poorest
fit (Table\,\ref{table:variability}). Likewise the scatter (standard
deviation) of the magnitudes of the source are significantly greater
than those of the comparison objects. For the two $V$-band light
curves, the scatter of the source and the comparisons are more
comparable, at least on MJD 52303, but the $\chi^{2}_{\nu}$ of the
constant fits to the source are again greater than those to the
comparison objects and inconsistent with being an acceptable fit.

While the $\chi^{2}_{\nu}$ of the constant fits to the $V$-band
comparison objects are all consistent, at $< 4\sigma$ level, with
those objects having constant magnitudes, the fits to the \ks-band
comparison stars are not consistent ($>5\sigma$) with that assumption.
\psf\ photometry, particularly in the NIR, is prone to underestimating
the actual errors on magnitude. This is due to the difficulty of
accurately modelling the \psf\ from sources in a crowded field (more
likely in NIR observations) and due to the difficulty of obtaining a
representative \psf\ from images where the width is not significantly
greater than 1 pixel, as is the case here.  Even if we normalise the
$\chi^{2}_{\nu}$ of the source by that of the worst case comparison,
we find that all the $V$- and \ks-band light curves are inconsistent
with a constant magnitude, though this should only be used as an
approximate guide.

If we use the maximum standard deviation of the comparison sources on
each epoch as an estimate of the ``background'' noise we can calculate
a corrected standard deviation of each source light curve and the root
mean square variability of the light curve, as a percentage of flux
(\%RMS, Table\,\ref{table:variability}). The error of the \%RMS of the
source is given as the \%RMS of the background and the $3\sigma$
upper limit to the variability is given as 3 times the background
level. The calculated \%RMSs imply that neither of the $V$-band light
curves display significant variability, while in all but one case, the
\ks\ light curves display variability of $\approx 20\%$. While this is
in contrast to the $\chi^{2}_{\nu}$ analysis of the light curves, it
is more robust as it is independent of any underestimate of the errors
on individual points (assuming that any underestimate is similar for
the source and for the comparison stars, which were chosen specifically 
to be of similar magnitude and hence, of a similar signal-to-noise ratio).

We used the {\small IRAF} task, {\tt pdm} --- an implementation of the
phase dispersion minimisation method of \cite{Stellingwerf1978:ApJ224}
--- to test if any of the variability of the light curves displayed a
periodicity.  However, all tests returned Stellingwerf statistics,
$\Theta \approx 1$ for all periods less than twice the duration of the
observations, implying no periodic variability.

\subsection{Spectral Energy Distributions}\label{section:results:sed}

In the following XSPEC fits, all 6 epochs of data are fit
 simultaneously, fixing the absorption ({\tt redden)} and
extinction ({\tt tbabs}), as well as other parameters on a case by
case basis, across all epochs.

Given that the observed quiescent magnitudes
\citep{Russell2011:MNRAS.416} or the limits placed on those magnitudes
by the 2MASS catalogue are magnitudes dimmer than we observe, we
assume that the companion star makes no significant impact on our
observed spectra. The NIR/optical data alone are poorly constrained
but consistent with a single reddened power law  ($F_{\nu}
\propto \nu^{\alpha}$), of different
normalisations, of spectral index $\alpha = -0.6 \pm 0.2$, and the
extinction $E_{B-V} = 1.2 \pm 0.1$, but these parameters are highly
degenerate and any values in the range $-1.4 \lesssim \alpha \lesssim
0.1$ for $0.7 \lesssim E_{B-V} \lesssim 1.7$ will give acceptable
fits. When we include the X-ray data a single power law is no longer
an acceptable fit, as the extrapolation of the NIR/optical spectral
index underestimates the X-ray flux.

Previous studies of the X-ray spectra of the 2002 outburst
\citep{Belloni2002:A&A.390} show that the system is in a hard state
with the spectra being well described by a power-law of spectral index
in the range of $-0.4$ to $-0.5$, without any thermal
component. However, studies of other black hole sources (e.g.,
\citealt{Miller2006:ApJ.652,Rykoff2007:ApJ.666,Reis2010:MNRAS.402,Reynolds2013:ApJ.769})
show that, even in the hard state, X-ray spectra can be fit by an
irradiated disk model that can also describe emission at optical
energies. We find that this model (implemented in XSPEC as {\tt
  diskir};
\citealt{Gierlinski2008:MNRAS.388,Gierlinski2009:MNRAS.392}) can well
describe the broadband data from NIR to X-ray. However, due to the
fact that the thermal component makes little contribution to the X-ray
flux above 3\,keV in this state, as demonstrated by the pure power law
models of \cite{Belloni2002:A&A.390}, the model is under-constrained
and many of the parameters related to the disk component cannot be
estimated with much certainty.  It is also true that the column
density is under-constrained due to the relative weakness of X-ray
absorption over the observed energies so we instead adopt that
measured from {\it Chandra} data
\citep{Tomsick2001:ApJ.563,Miller2003:MNRAS.338}.  The extinction is
set to the value of $E_{B-V} = 1.02 \pm 0.05$, which is derived from a
fit to the only epoch (MJD 52293/4) that includes both NIR and optical
data which may best constrain extinction.  This is consistent with
that implied, via the relationship of \cite{Guver2009:MNRAS.400}, from
the X-ray absorption of this source.

The fits to the irradiated disk model are poorly constrained but
return photon indices of the power law component, $\Gamma \sim1.5$,
and disk temperatures, $kT_{{\rm disk}} \sim0.2$\,keV, at all epochs.
These values are in agreement with the photon indices derived for this
source by \cite{Belloni2002:A&A.390} and the disk temperatures
measured in the hard state of other \lmxbs\ (e.g.,
\citealt{Miller2006:ApJ.652,Rykoff2007:ApJ.666}).
The other parameters of the fit, even when fixed across epochs, are
unconstrained but are in broad agreement with the underlying
parameters of \cite{Gierlinski2009:MNRAS.392}.  Given the poorly
constrained nature of these spectra it is important to caution that
the fit results should not be over interpreted, however we can state
that the data are at least consistent with the irradiated disk model
for a range of realistic, physical parameters and energies.


\section{Discussion}\label{section:discussion}

X-ray observations of the 2001 and 2002 outbursts of \xtefifteen\
\citep{Tomsick2001:IAUC.7575,Belloni2002:A&A.390} suggest that the
source was in a hard state only, with no reports of a transition to a
soft or intermediate state. Defining the state of an \lmxb\ from NIR/optical
observations is not as straight forward as from the X-ray due to the
multiple emission mechanisms -- such as the accretion disk, radio jet,
corona, reprocessing (see section \ref{section:intro}) -- that
contribute at those wavelengths. However, variability, which is
attributed to the non-thermal emission of the jet or corona, has been
observed in the hard state, at least at NIR wavelengths (e.g.,
\citealt{Casella2010:MNRAS.404,Chaty2011:A&A.529}). In the past
decade, a number of correlations have also been suggested that
indicate which state the system is in without the need to know exactly
which emission mechanism is contributing to the fluxes.  This method
uses the observed correlations between the X-ray luminosities and the
NIR/optical \citep{Russell2006:MNRAS.371} or radio (e.g.,
\citealt{Corbel2000:A&A.359,Corbel2003:A&A.400,Gallo2003:MNRAS.344,Fender2010:MNRAS.406,Coriat2011:MNRAS.414})
luminosities in different states to imply which state the system is
in.

Our detections of short term variability in the high-cadence \ks-band
light curves of $\approx 20\%$ and non-detections in the $V$-band is
consistent with the variabilities implied during the failed 2003
outburst of this source \citep{Chaty2011:A&A.529}. In that outburst,
over a similar range of frequencies to ours ($\approx 10^{-4} -
10^{-1}$ Hz), \ks-band variability was detected at a level of $7.2 \pm
2.2 \%$ in contrast to a $V$-band upper limit of $< 28.3\%$.  These
imply that the radio jet, or perhaps the high-energy corona, are
making a significant contribution to the NIR flux. Compact radio jets
are only observed in the hard state and while the corona can
contribute in the soft state, it is usually weak, so any significant
emission that can be associated with either implies that the source is
in a hard state. Radio emission, consistent with optically thick
emission from a compact jet, was observed from this source during the
2002 outburst \citep{Corbel2002:IAUC.7795} and, given its flux of
$\approx$2.5\,mJy and spectral index of $0.07 \pm 0.11$, it is
plausible that it contributed to the NIR flux. Unfortunately, while
further radio observations of this source have been obtained with the
same instrument during the 2001, 2002, and 2003 outbursts, they have
yet to be published and it is beyond the scope of this paper to do so.
Assuming that the observed variability is due to the radio jet implies
that the jet's spectral break frequency is at NIR wavelengths -- a
result consistent with the previously implied break frequencies, both
for this source in full outburst, and other \lmxb\ systems
\citep{Russell2013:MNRAS.429}.

The featureless spectra indicate that there is little direct emission
from the accretion disk which would be expected in the soft
state. While our SEDs are not well constrained, they are consistent
(see section \ref{section:results:sed}) with the flux being due to
reprocessing of X-rays in a relatively cool ($\sim 0.2$\,keV)
accretion disk, as expected in the hard state ( e.g.,
\citealt{Miller2006:ApJ.652,Rykoff2007:ApJ.666}; a much higher disk
temperature of $\sim 1$\,keV is expected in the soft state, e.g.,
\citealt{Sobczak2000:ApJ.544}).
The absolute $V$-band magnitudes at the various epochs ($M_{V} \approx
1-2$, assuming a distance of $5.3 \pm 2.3$ kpc;
\citealt{Jonker2004:MNRAS.354}) are also consistent with observed
correlation with $\Sigma = (L_{{\rm X}} / L_{{\rm Edd}} )^{1/2}
P^{2/3} \approx -0.3$
\citep{vanParadijs1994:A&A.290,Deutsch2000:ApJ.530} if we use the
observed period, $P = 1.541 0\pm 0.009$ days \citep{Jain2001:ApJ546},
and an estimated mass of $ \simeq 7-10 {\rm M}_{\odot}$
\citep{Orosz2002:ApJ.568,Munoz2008:MNRAS.385}. This agreement of the
observed magnitudes with this relationship is also consistent with
those magnitudes being due to reprocessing.

Comparing our derived NIR/optical and X-ray luminosities (along with
those from the failed 2003 outburst;
\citealt{Arefev2004:AstL.30,Chaty2011:A&A.529}) with the observed
values for other \lmxbs\ \citep{Russell2006:MNRAS.371} we find no
significant deviation from the hard state correlation. It has been
shown \citep{Curran2012:A&A.547} that sources deviate from this
correlation early in the intermediate state so this agreement adds
further weight to the suggestion that the system was in the hard state
at the time of the optical observations.
During the 2001 outburst, our observations span most of the X-ray
activity (see figure \ref{fig.rxte}) without displaying any evidence
of reaching a hard or intermediate state. Observations of the 2002
outburst and the single epoch of observations in the 2003 outburst 
\citep{Chaty2011:A&A.529} were obtained only after the X-ray flux had
already peaked and hence we cannot rule out that a transition to an
intermediate state occurred; however, if a soft or intermediate state was
reached, the transition back to the hard state would not be expected
until late times when the accretion rate had dropped significantly.


\section{Conclusions}\label{section:conclusions}

The NIR/optical data of the black hole \lmxb\ system, \xtefifteen,
while being consistent with having originated from reprocessing of
X-rays in the accretion disk, display variability indicative of a
contribution from the radio jet at NIR wavelengths. Hence, the
NIR/optical likely combines both emission from the jet and
reprocessing. The contribution of the radio jet at such high
frequencies is consistent with both previous observations of this
source in full outburst and other \lmxb\ systems
\citep{Russell2013:MNRAS.429}, and supports the hard state
classification of the system at the time of observations. A comparison
of the NIR/optical and X-ray luminosities with those of other
\lmxbs\ displays no deviation from the observed hard state
correlations \citep{Russell2006:MNRAS.371}. This suggests that the
failed outbursts of 2001, 2002, and 2003 did not transition to an
intermediate state, or display signs of jet quenching, but remained in
a true, hard state throughout the outburst.
Failed outbursts seem only to differ from standard outbursts by their
failure to quench the radio jet and reach accretion disk dominated
emission and not in their underlying, time-independent, physical
structure.  Studying them at multiple wavelengths -- to constrain
physical parameters, such as the frequency of the jet break, the
accretion disk temperature and radius -- is required to reveal how jet
suppression and reactivation relates to accretion parameters in both
failed and successful outbursts of \lmxbs.

\begin{acknowledgements}  
  We thank the anonymous referee for their useful comments. This work
  was supported by the Australian Research Council's Discovery
  Projects funding scheme (project number DP120102393) and by the
  Centre National d'Etudes Spatiales (CNES). This work is based on
  observations obtained with MINE: the Multi-wavelength INTEGRAL
  NEtwork.  This research has made use of NASA's Astrophysics Data
  System, the SIMBAD database, operated at CDS, Strasbourg, France and
  quick-look results provided by the ASM/RXTE team.
\end{acknowledgements}

\end{document}